\documentclass[aps,prc,twocolumn,groupedaddress,showpacs]{revtex4}
\usepackage{graphicx}
\usepackage[rightcaption]{sidecap}
\begin{document}
\newcommand{\vdot}{\mbox{\boldmath $\cdot$}}
\newcommand{\vect}[1]{\mbox{${\bf #1}$}}
\newcommand{\beq}{\begin{equation}}
\newcommand{\eeq}{\end{equation}}
\newcommand{\etal}{ \emph{et al\/}}
\newcommand{\gvect}[1]{\mbox{\boldmath$ #1$}}
\def\lbar{\mathchar'26\mkern-10mu\lambda}
\newcommand{\hlf}{\mbox{$\frac{1}{2}$}}
\newcommand{\aS}{\mbox{$\arg{S_L}$}\,\,}
\newcommand{\mS}{\mbox{$|S_L|$}\,\,}
\newcommand{\phm}{\phantom{0}}
\newcommand{\ri}{\mbox{${\rm i}$}}
\newcommand{\rri}{\text{i}}
\def\nuc#1#2{\relax\ifmmode{}^{#1}{\protect\text{#2}}\else${}^{#1}$#2\fi}

\title{Emergence of a secondary rainbow and the dynamical polarization potential for \nuc{16}{O} on \nuc{12}{C} at 330 MeV}
\author{R. S. Mackintosh}
\email{raymond.mackintosh@open.ac.uk}
\affiliation{Department of Physical Sciences, The Open University, Milton Keynes, MK7 6AA, UK}
\author{Y. Hirabayashi}
\email{hirabay@iic.hokudai.ac.jp}
\affiliation{Information Initiative Center, Hokkaido University, Sapporo 060-0811, Japan}
\author{S. Ohkubo}
\email{ohkubo@yukawa.kyoto-u.ac.jp}
\affiliation{Research Center for Nuclear Physics, Osaka University, Ibaraki, Osaka 567-0047,
 Japan}
\affiliation{University of Kochi,  Kochi 780-8515, Japan }

\date{draft of \today}

\begin{abstract}
\begin{description}

\item{Background:} It was shown recently that an anomaly in the elastic scattering of 
\nuc{16}{O} on \nuc{12}{C}  at around 300 MeV is resolved by including within the
scattering model the inelastic excitation of specific collective excitations of both nuclei,
leading to a secondary rainbow.  There is very little systematic knowledge concerning 
the contribution of collective excitations to the interaction between nuclei, particularly
in the overlap region when neither interacting nuclei are light nuclei.
\item{Purpose:} To study the dynamic polarization potential (DPP) generated by channel
coupling that has been experimentally validated for a case (\nuc{16}{O} on \nuc{12}{C}
at around 300 MeV) where scattering is sensitive to the nuclear potential over a wide
radial range; to exhibit evidence of the non-locality due to collective coupling;
to validate, or otherwise invalidate, the representation of the DPP by uniform renormalizing
folding models or global potentials.
\item{Methods:} S-matrix to potential, $S_{L} \rightarrow V(r) $, inversion yields local
potentials that reproduce the elastic channel S-matrix of coupled channel calculations.
Subtracting the elastic channel uncoupled potential yields a local $L$-independent 
representation of the DPP. The dependence of the DPP on the nature of the coupled states 
and other parameters can be studied. 
\item{Results:}  Local DPPs were found due to the excitation of \nuc{12}{C}  and  the 
combined excitation of \nuc{16}{O} and \nuc{12}{C} . The radial forms were different for
the two cases, but each were very different from a uniform renormalization of the potential.
The full coupling led to a 10\% increase in the volume integral of the real potential.
Evidence for the non-locality of the 
underlying formal DPP and for the effect of direct coupling between the collective states
was presented, 
\item{Conclusions:} The local DPP generating the secondary rainbow has been identified. 
In general, DPPs have forms that depend on the nature of the specific
excitations generating it, but as in this case, they cannot be represented by a uniform 
renormalization of a global model or folding model potential. The method employed
herein is a useful tool for further exploration of  the contribution of collective 
excitations to internuclear potentials concerning which there is still remarkably little general information.

\end{description}
\end{abstract}


\maketitle

\par
\section{INTRODUCTION}
The collective excitation of interacting nuclei strongly influences the elastic scattering 
between those nuclei. This was clearly demonstrated in Ref.~\cite{OH1}  by the discovery of 
a secondary rainbow in \nuc{16}{O} on \nuc{12}{C}  elastic scattering, resolving a significant
anomalous situation involving the elastic scattering of this pair of nuclei. A global
model that had proven satisfactory for energies from 62 to 1503 MeV was found to be 
significantly inadequate when confronted with wide angular range data  at $E_L = 281$ MeV.
At this and similar energies, the Airy minimum of nuclear rainbow scattering appears at a
much larger angle, $\theta \sim 70^{\circ}$, than expected with the global model. 
As described in Ref.~\cite{OH1}, this problem was resolved by including excitation 
of the $2^+$ and $3^-$ states of \nuc{12}{C} and the $3^-$ and $2^+$ states of \nuc{16}{O}
within an extended folding model. Within this model, a secondary nuclear rainbow appears 
having an Airy minimum at a large angle, consistent with experiment over a range of
energies, with the primary rainbow at more forward angles being somewhat obscured. 
In this way, including the excitation of the various collective states resolves a 
significant anomaly and leads to an understanding of further phenomena related to nuclear
rainbow scattering~\cite{OH2}.

The excitation of collective states plays an important role in the dynamics of the
interactions between all pairs of nuclei. However, there is little systematic knowledge,
particularly for heavy ion scattering, concerning the contribution of such collective
states to the scattering potential.  In this paper we shall demonstrate a means of mitigating this lack of information. The case we study, \nuc{16}{O}  on \nuc{12}{C} at 330 MeV, is one in which the contribution of collective states has been shown to resolve a known anomaly. The results will throw light on the limitations of standard phenomenology that is based on parameterized forms or based on folding models that lead to smooth potentials.  They also suggest a richness of phenomena concerning the dynamics of interacting nuclei, including evidence for dynamically generated non-locality.  The procedure exemplified in this paper is of wide applicability. Section~\ref{dpp} reviews relevant general properties of the DPP, Section~\ref{model} introduces the model and discusses
the DPP due to excitations of \nuc{12}{C}, Section~\ref{oxy} presents the effects of excitations of
\nuc{16}{O}, Section~\ref{disco} discusses the results and the implications, and  Section~\ref{summa}
briefly summarizes.

\section{THE DYNAMIC POLARIZATION POTENTIAL}\label{dpp}
Within  a potential model, the effects of channel coupling can be represented as a dynamic polarization potential, DPP,  added  to a folding model potential, see Ref.~\cite{feshbach, satchler} for example. The formal DPP  is both $L$-dependent and non-local, the non-locality being in addition to that which arises from exchange processes. Adding such a non-local, $L$-dependent potential to the folding model potential yields a potential that is itself non-local and $L$-dependent. Such a potential is not easily comparable with local phenomenological potentials, see for example Ref.~\cite{rawit87}. However it is always possible, using  $S_L \rightarrow V(r)$ inversion,  to find a local and  $L$-independent potential that  yields the same S-matrix, and hence all elastic scattering observables, as any non-local and $L$-dependent potential.  In this way, we can determine a potential that exactly reproduces the $S$-matrix of the sum of the formal DPP and the folding model potential.  This local  potential  can be compared with potentials determined by means of precision phenomenological fitting.  We shall generally refer to the local DPP  found by $S_L \rightarrow V(r)$  inversion of the coupled channel $S_L$, followed by subtraction of the folding model (`bare') potential,  as `the DPP'. However, when it is relevant that the true, formal DPP~\cite{feshbach,satchler}, is non-local, we refer to this as the `underlying DPP'. We shall present some evidence for that non-locality.  

For the particular case of proton scattering from nuclei, it has been shown~\cite{mk90} that 
collective excitations give rise to local DPPs that are strongly undulatory. This may be related
to the non-locality and/or $L$-dependence of the underlying formal DPP; specific evidence
concerning this non-locality for proton scattering was presented in 
Ref.~\cite{NM14}.  Less is known about the properties of the DPP for multi-nucleon projectiles 
than for nucleons, apart from a few studies for lighter heavy ions  and various cases of DPPs
generated by projectile breakup.   The effects of channel coupling are commonly absorbed 
into an overall renormalization of a global optical model potential (OMP) or folding model
(FM) potential, a practice that we shall comment upon later. This procedure may be
justified for those heavy ion reactions for which only the nuclear surface region is 
significant.

\section{THE MODEL AND DPP FOR  \nuc{16}{O}+\nuc{12}{C} SCATTERING}\label{model}
The case of scattering studied here, \nuc{16}{O} on \nuc{12}{C}  at some hundreds of MeV,
is characterized by a high degree of penetration of the two nuclei, so there is some 
sensitivity at large angles to the potential for almost complete overlap of the two nuclei.
This motivates the determination of the DPP over a wide radial range. The coupled channel
model of Ref.~\cite{OH1} is much more realistic than earlier models for which DPPs for
the same pair of nuclei were calculated~\cite{mc89,mc93}, and is validated by its detailed
fit to wide angular range  elastic scattering data, solving the problem of the rainbow
scattering angle.

For 330 MeV incident energy, a standard double folding (DF) model calculation, with no
inelastic coupling, failed to fit beyond about 60$^\circ$ elastic scattering differential
cross section data extending  beyond 90$^\circ$. The initial extended double folding (EDF)
coupled channel calculation included coupling to the collective $2^+$ and $3^-$ states of \nuc{12}{C}  at 4.44 and 9.64 MeV respectively. A further calculation (`full' ) also included  the 
$3^-$ and $2^+$ states of \nuc{16}{O} at 6.13 and 6.92 MeV.  A Woods Saxon imaginary term 
is added to the real DF and EDF interactions. This is shallower for the initial EDF 
calculations with a further reduction in depth and diffusivity for the full calculation. The DPPs that we present for the initial and full cases are calculated by subtracting the relevant bare (diagonal, uncoupled) potential, with its appropriate imaginary term,  from the potential that is found by inversion to reproduce the EDF model elastic channel $S$-matrix. Since the DPP for any given case is  somewhat dependent on the imaginary part of the bare potential, this should be borne in mind when comparing the DPPs for the initial and full cases.

\begin{figure} [tbh]
\includegraphics[keepaspectratio,width=7.0cm] {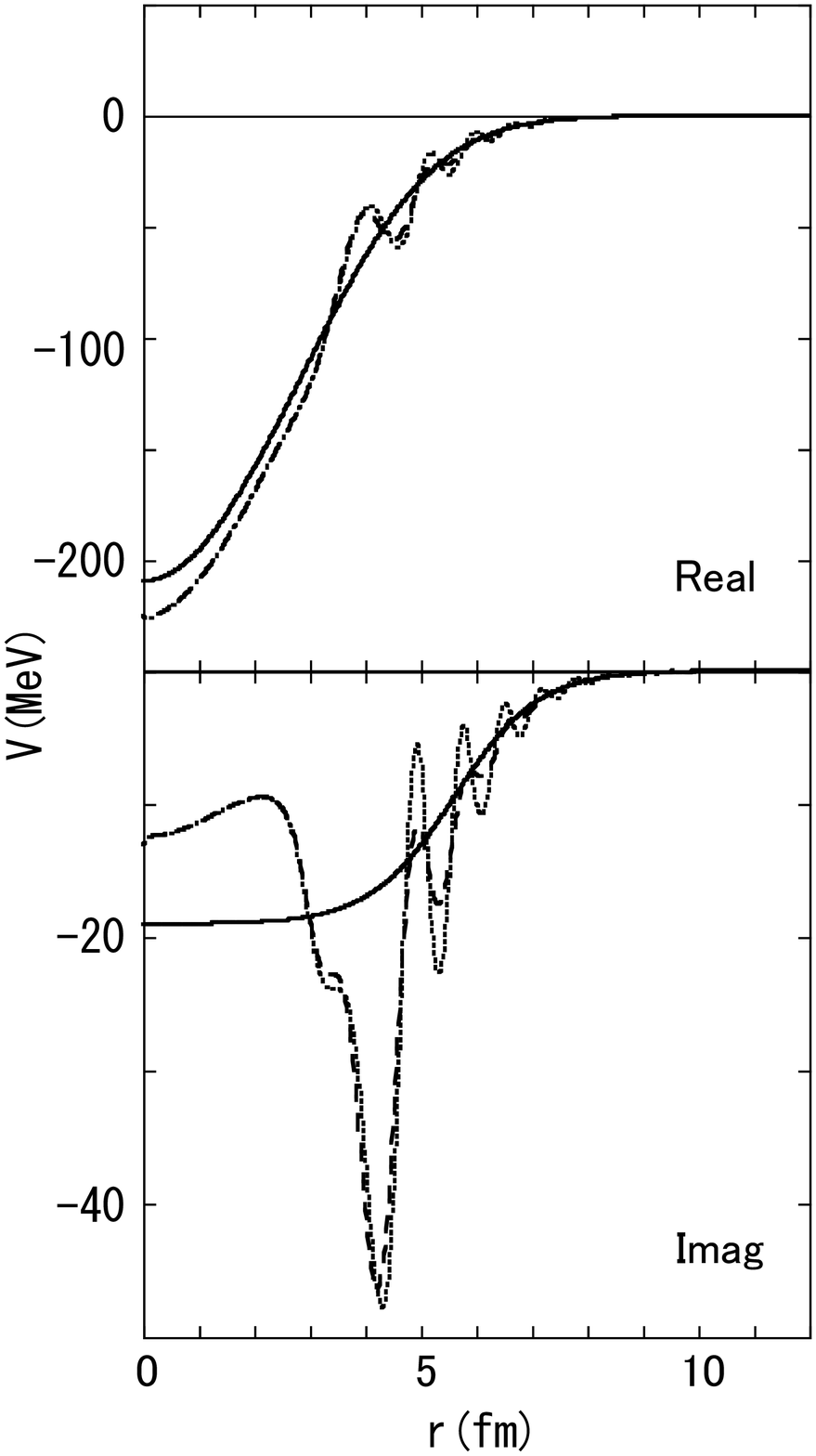}
 \protect\caption{\label{dpp9} {
For 330 MeV \nuc{16}{O} scattering from $^{12}$C,
the inverted potential fitting $S_L$  in the presence of coupling to two states of $^{12}$C. 
The solid line  is for the  bare potential (no coupling)  and the more
oscillatory dotted line is the inverted potential. The dashed line represents 
the potential at an earlier stage of the iterative inversion. } }
\end{figure}

\begin{figure} [b]
\includegraphics[keepaspectratio,width=7.0cm] {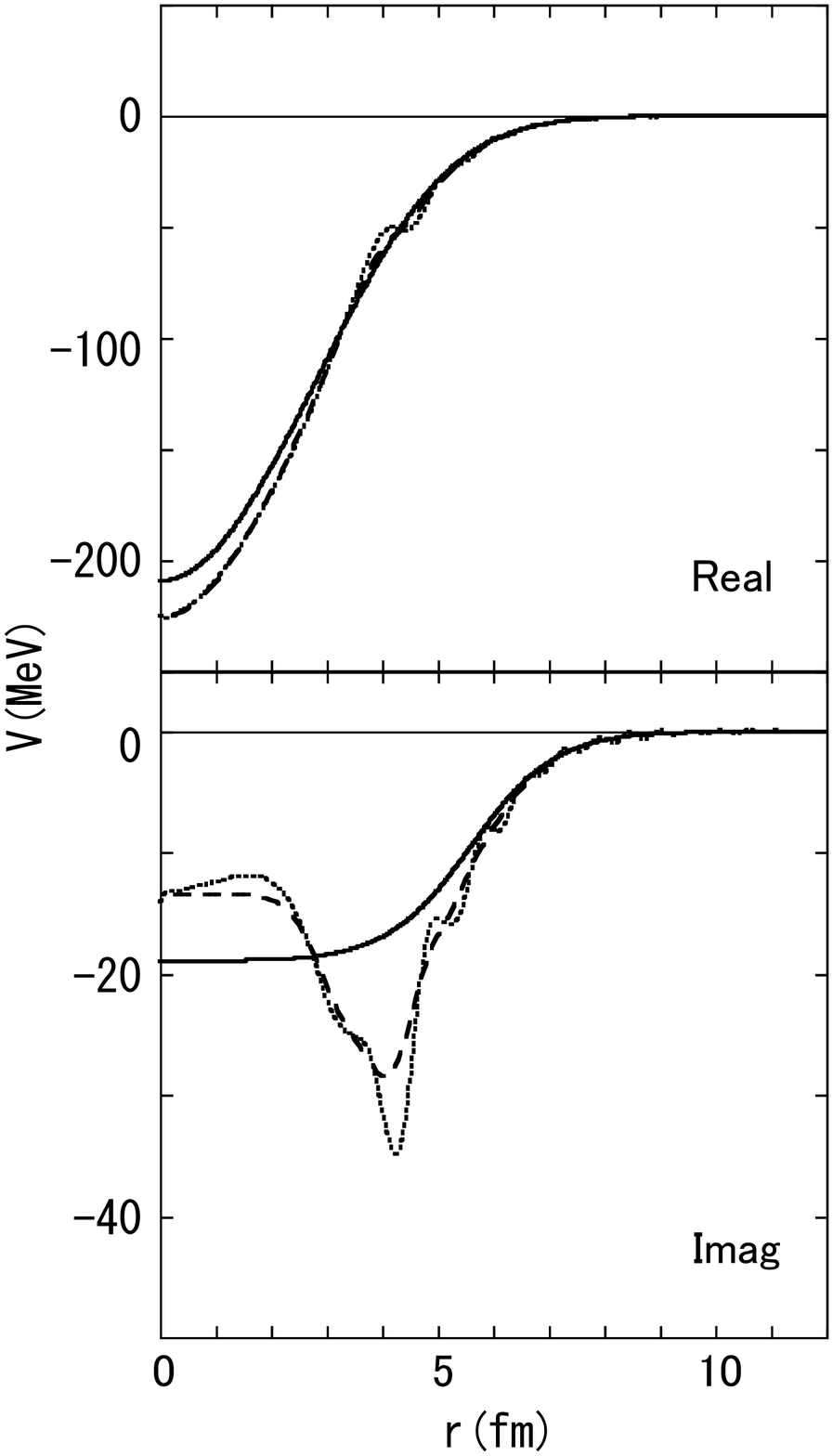}
\protect\caption{\label{nmc} {
The dotted line shows the inverted potential for coupling to $2^+$ and $3^-$ states 
of \nuc{12}{C} with no direct coupling between these states. The dashed line is the inverted potential 
for coupling to just the $2^+$ state of \nuc{12}{C}. The solid line is the  bare potential (no coupling). }}
\end{figure}

We calculated the DPPs for the following four sets of  inelastic excitations of \nuc{12}{C},
all involving the same initial bare imaginary term: (i) the $2^+$ state, (ii) the $3^-$ 
state, (iii) the $2^+$ and $3^-$ states with direct coupling between these excited states,
and (iv) the $2^+$ and $3^-$ states with no direct coupling between them.  We also
calculated the DPPs for the full case including the $2^+$ and $3^-$ states of \nuc{16}{O}
with the relevant imaginary bare potential. In no case was the imaginary potential deformed,
so the coupling potential was purely real.

\begin{table*}[t]
\caption{For \nuc{16}{O} scattering from \nuc{12}{C} at 330 MeV, the
characteristics of the DPP due to the coupling specified in column 1, 
for the first 5 rows for \nuc{12}{C} states only. The columns $\Delta J_{\rm R}$ and
 $\Delta J_{\rm I}$ give the change in the volume  integral (per nucleon pair)
 of the real and imaginary components of the DPP induced by the coupling. 
 The columns  $\Delta R_{\rm R}(\rm rms)$ and $\Delta R_{\rm I}(\rm rms)$ 
 respectively give the change in rms radius of the real and imaginary central 
components.  Negative $\Delta J_{\rm R}$ corresponds to repulsion. 
In line 5, `ndc' indicates no direct coupling between the $2^+$ and $3^-$ states. 
The two final columns present, respectively, the change in the total reaction cross
 section induced by the coupling, and the integrated cross section to the 
specific coupled reaction channels.}
\begin{tabular}{lrrrrrr} 
\hline
Coupling &$\Delta J_{\rm R}$ (MeV fm$^3)$ &$\Delta R_{\rm R}(\rm rms) $ (fm)& $\Delta J_{\rm I}$
(MeV fm$^3)$ &   $\Delta R_{\rm I}(\rm rms)$ (fm)& $\Delta$ CS (mb) &Inel.\ CS (mb)\\ \hline\hline
$2^+$                  & 7.67 & $-0.0077$ & 20.68& $-0.0815$ &11.8 & 11.4\\
$3^-$                   & $-0.34$ & 0.0036 & 2.37 & $-0.0007$ & 4.3 & 3.9 \\
$2^+$ and $3^-$& 5.85 & 0.0013       & 28.56& $-0.0917$ & 15.6& 14.8 \\ \hline
$\Sigma 2^+ 3^-$& 7.33& $-0.0041$& 23.06 & $ -0.0822$& 16.1& 15.3 \\ \hline
$2^+$and $3^-$ ndc& 6.12& $-0.0019$& 25.21 & $ -0.0909$& 15.7& 14.9 \\ \hline
Full coupling& 28.74 & -0.0064 &  35.96 & 0.0106 & 70.4 & 57.4\\ \hline 
\end{tabular} 
\label{dpps}
\end{table*}

The elastic scattering $S$-matrix from the coupled channel calculation was subjected
to $S_L \rightarrow V(r)$ inversion using the iterative-perturbative, IP, inversion
procedure~\cite{MK82,ip2,kukmac,arxiv,spedia}. The resulting potential exactly reproduces 
the elastic scattering from the coupled channel, CC, calculation. Subtracting the diagonal
elastic channel potential from the inverted potential yields a local $L$-independent 
representation of the DPP corresponding to whatever particular channels inelastic channels
were included in the CC calculation.

It is useful to quantify the DPPs for the various cases in a simple way
by subtracting from the volume integrals and rms radii for the inverted potential the same
quantities for the bare potentials. This  gives the changes that are induced by the coupling. The
results are presented in Table~\ref{dpps}. Specifically, we calculate
$\Delta J_{\rm R}$, $\Delta R_{\rm R}(\rm rms)$, $\Delta J_{\rm I}$ and
 $\Delta R_{\rm I}(\rm rms)$ which are the changes in the real volume integral
and rms radius and the imaginary volume integral and rms radius.  All volume integrals 
are conventionally defined in terms of nucleon pairs~\cite{satchler}. 
The  quantities $\Delta J_{\rm R}$ and $\Delta J_{\rm I}$ are the volume integrals of the 
real and imaginary parts of the DPP while the interpretation of the radial quantities is 
less direct. The quantity $\Delta J_{\rm R}$ alone gives an incomplete view of the change
in the real potential since this typically has both attractive and repulsive regions.
However since $J_{\rm R}$ for the bare potential is 278.48 MeV fm$^3$, it can be seen 
that for the full case, the attraction due to the coupling corresponds to a 10\% increase in volume integral; 
in fact there is a 30\% increase around 3.5 fm, see Fig.~\ref{full} discussed 
in Section~\ref{oxy} below.

\begin{figure}[t]
\includegraphics[keepaspectratio,width=7.0cm] {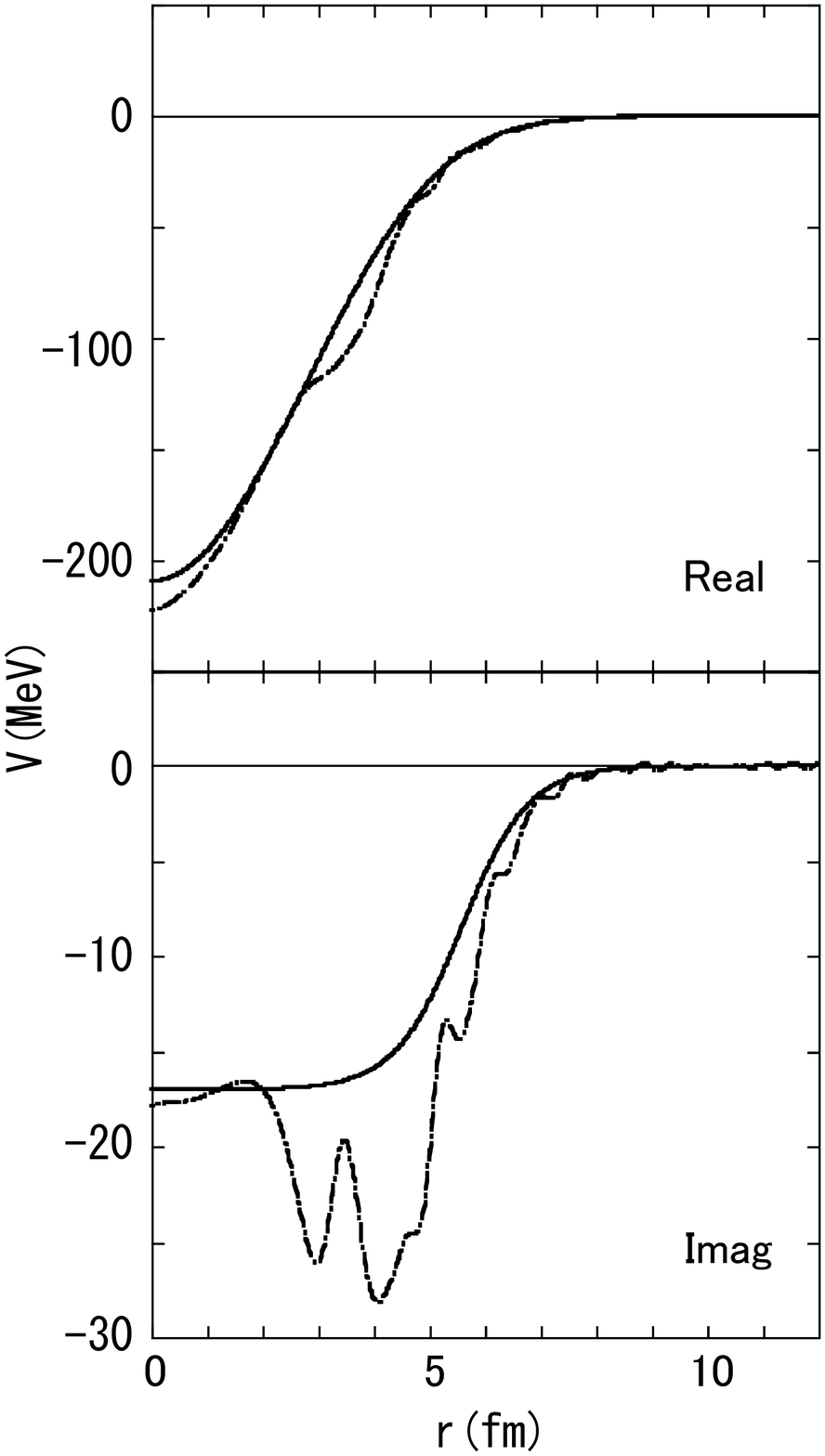}
 \protect\caption{\label{full} {
For 330 MeV \nuc{16}{O} scattering from $^{12}$C,
the inverted potential fitting $S_L$  for the full coupling to states of $^{12}$C 
and $^{16}$O. The solid line is for the  bare potential (no coupling).  The vertical scale
for the imaginary part is different from the scale used in Fig.~\ref{dpp9} and
Fig.~\ref{nmc}. The different imaginary bare potential for this case can be seen. } }
\end{figure}

\begin{figure} [tbh]
\includegraphics[keepaspectratio,width=7.0cm] {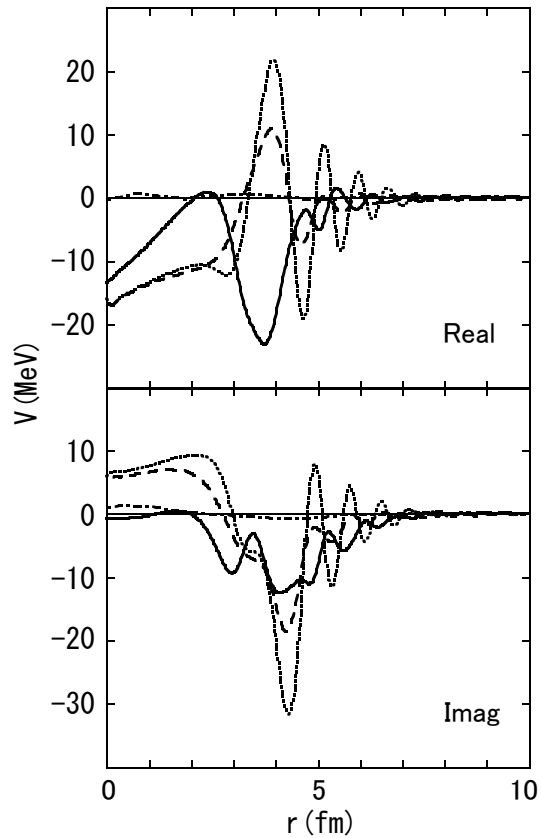}
 \protect\caption{\label{3dpps} { The DPPs implicit in Fig.~\ref{dpp9}, Fig.~\ref{nmc} and Fig.~\ref{full} are 
compared directly. The dotted line is the DPP for the excitation of both states of \nuc{12}{C} with direct coupling between them, the dashed line is the DPP when the direct coupling between the states of \nuc{12}{C}  is omitted, and the solid line  is the DPP for the `full' case, i.e.\ when the excitation of states of \nuc{16}{O} is added. The dash-dot line is the small magnitude DPP due to the $3^-$ state of \nuc{12}{C} alone. } }
\end{figure}

\begin{figure} [t]
\includegraphics[keepaspectratio,width=7.0cm] {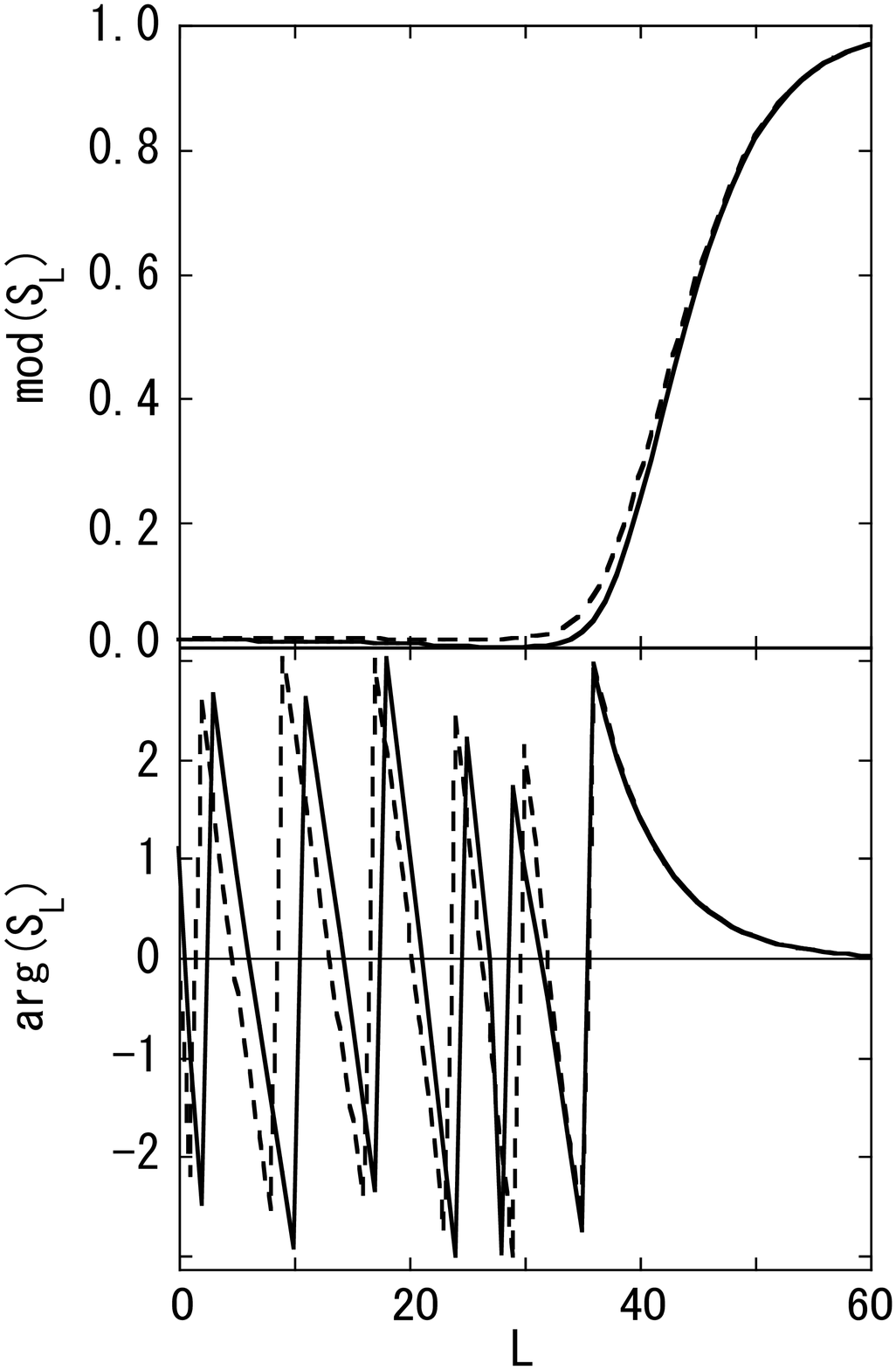}
 \protect\caption{\label{SL60} {
 Change in the elastic scattering $S$-matrix $S_L$ for  coupling to the $2^+$ and $3^-$ states 
of \nuc{12}{C}.  The broken line is for  the bare potential with no coupling, the solid line is with 
coupling.  Upper panel is  $|S_L|$, lower panel is $\arg S_L$. The apparent discontinuity 
in $\arg S_L$ reflects the principal value of arctan, $-\pi \le \arg S_L \le \pi$} }
\end{figure}

Insight into the nonlocality of the underlying formal DPP can be gained from a comparison 
of lines 4 and 5 of Table~\ref{dpps} based on the fact that the local equivalent of the
sum of two non-local potentials is not the sum of the local equivalents of each potential.
The non-local DPPs due to the excitation of two states when there is no direct coupling 
between those states add to give the non-local DPP due to those states. However, the
presence of non-locality is indicated by the non-exact addition of the local equivalent
DPPs. Line 4 of Table~\ref{dpps} gives the numerical sum of the numbers in rows 1 and 2,
and we draw attention to the volume integrals which are clearly unequal to the volume 
integrals in line 5, the case in which the two states are excited with no direct coupling
between them. Although not directly related to the issue of underlying non-locality, it is
also of interest to compare lines 3 and 5; it appears that  direct coupling between the
states somewhat reduces the attractive effect but somewhat increases the absorptive effect
of the coupling, with no significant effect on the total inelastic or reaction cross sections.

\begin{figure} [tbh]
\includegraphics[keepaspectratio,width=7.0cm] {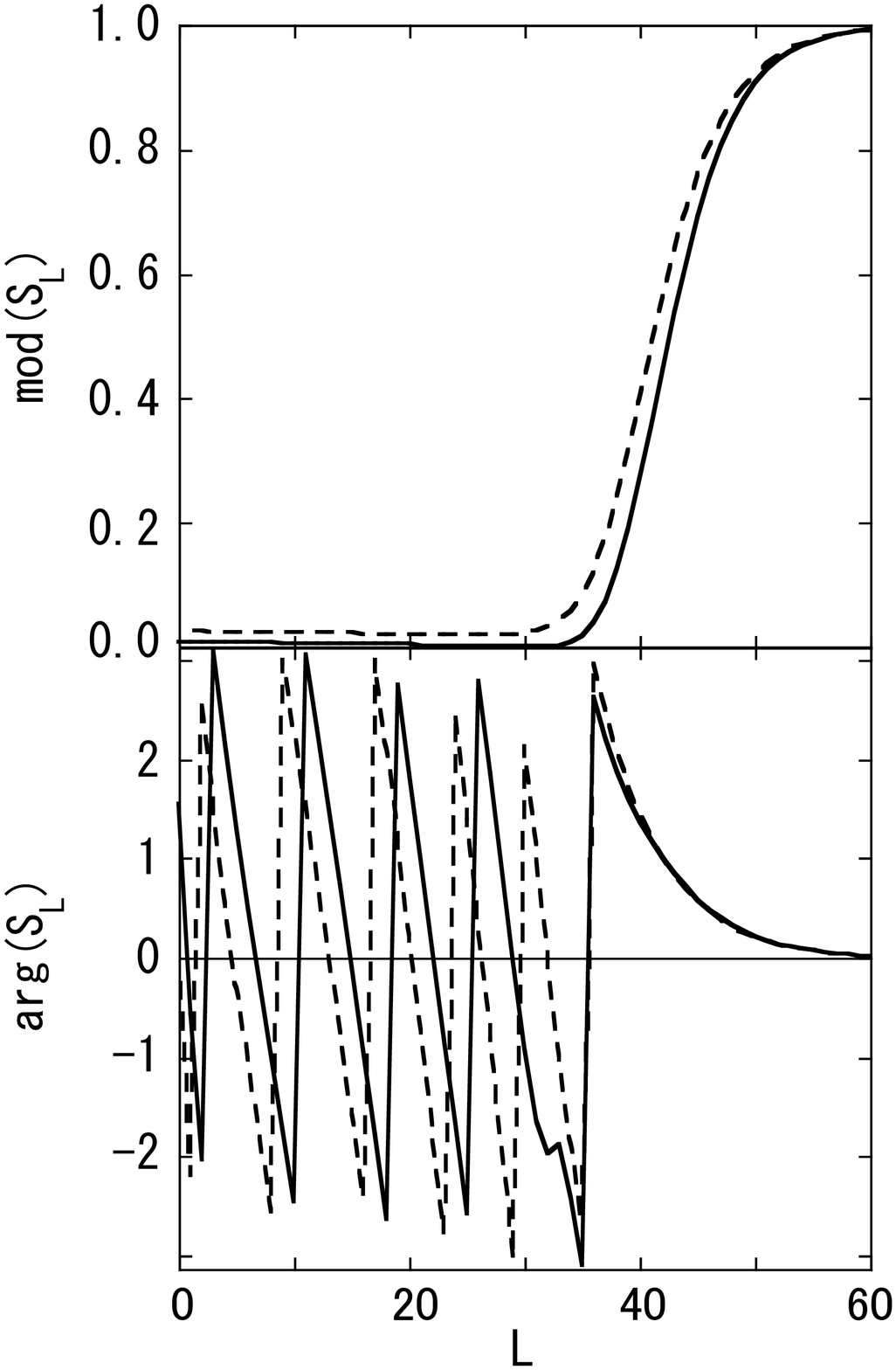}
 \protect\caption{\label{sm0full} { 
Change in $S_L$ for full calculations. The broken line is for the bare potential 
with no coupling, the solid line is for coupling to the states of \nuc{12}{C}
 and \nuc{16}{O}.   $|S_L|$ is in the upper panel and $\arg S_L$ in the lower panel.} }
\end{figure}

The bare and inverted potentials for the case with coupling to the  $2^+$ and $3^-$ states,
with direct coupling between them, are shown in Fig.~\ref{dpp9}.  The dotted line corresponds
to the potential quantified in line 3 of Table~\ref{dpps} and the dashed line corresponds
to the slightly less exact fit to $S_L$ of an earlier iteration of the inversion. The
development of quite sharp undulations appears to be genuine, and we comment on it later; 
the volume integrals and other characteristics of the two potentials are quite close. 
The DPP can be read off as the difference between the dotted line and the solid line
representing the bare potential. The attractive nature of the real DPP for $r \le 3.5$ fm,
and the emissive character of the imaginary DPP for $r \le 3$ fm, are well determined by
the inversion, and are within the radial range within which the scattering out to 
$90^\circ$ is sensitive, as verified with notch tests. It was found that somewhat 
less undulatory potentials could be found if a small Majorana ($(-1)^L$) term was 
included in the inversion, but this might have been simulating some other form of 
$L$-dependence, and this requires further investigation. 

The real and imaginary DPPs when the direct coupling between the  $2^+$ and $3^-$ states 
is turned off are similar but somewhat smoother, as can be seen from Fig.~\ref{nmc}. In this
case it was readily possible to get a very close reproduction of $S_L$ without the sharp 
undulations seen in Fig.~\ref{dpp9}. It was possible to get a slightly smoother potential
by stopping the inversion process with fewer iterations, but the undulations are clearly
less in amplitude than those seen in Fig.~\ref{dpp9}. It is possible that direct coupling between 
the  states generates some $L$-dependence or generates interfering amplitudes of some kind.
The undulations disappear when the $3^-$ state excitation is omitted, as seen in the dashed line
in Fig.~\ref{nmc} which shows the DPP for the excitation of just the $2^+$ state of \nuc{12}{C}.
The undulations in the dotted line clearly arise from an interference between amplitudes, the 
amplitude corresponding to the $3^-$ state being very small.  This can be seen from the small magnitude 
of the DPP for coupling to the $3^-$ state alone, which will be shown in  Fig.~\ref{3dpps} discussed below.

A comparison of the first three lines of Table~\ref{dpps} suggests that the $3^-$ state contributes 
much less to the DPP than the $2^+$ state. Although the $3^-$ state  does not contribute to the generation
of the secondary rainbow, as seen in Fig. 2 of Ref.~\cite{OH1}, it does make  a small but
non-negligable contribution to the angular distribution.  This is consistent with its undularity  
effect on the radial shape of the DPP, as noted above. Although the coupling to the $3^-$ state 
generates clear interference effects, it is the coupling to the $2^+$ state that is the key to generating 
the secondary rainbow.

Some calculations were also carried out at 300 MeV incident energy. We noted above that, for 330 MeV 
incident energy, the DPP resulting from the simultaneous excitation of the $2^+$ and $3^-$ states of 
$^{12}$C was more undulatory when there was direct coupling between these two excitations than when
 there was no such direct coupling. This is not an artifact of the inversion process; the same increased undulatory character is also observed at an incident energy of 300 MeV when direct coupling is included between the excited state channels. The DPPs themselves at 300 MeV  had the same general character as for 330 MeV and the direct coupling between excitations had almost no effect on the total reaction cross section, as was also the case at 330 MeV. Direct coupling between inelastic  excitations may therefore be a source of undulations in nucleus-nucleus interactions in general, although the mechanism is obscure at present. This deserves both theoretical and phenomenological study.

\section{EXCITATION OF \nuc{16}{O}}\label{oxy}
Although instructive and crucial for the effects reported in Ref.~\cite{OH1}, the DPPs
due to the excitation of the two states of \nuc{12}{C} are not the whole story.
For the `full' case, the excitation of \nuc{16}{O} substantially modifies the DPPs, 
see Fig.~\ref{full}: the real part now has substantial attraction around 4 fm, leading 
to the 10\% increase in volume integral noted above, and the imaginary part loses almost
all the emissivity within 3 fm, and is somewhat deeper towards the nuclear surface.  The imaginary bare
potential is shallower in the full case than in the other cases. This can easily be seen in Fig.~\ref{full}. 
The imaginary bare potential also had smaller diffusivity, as can be seen with closer inspection of this figure.

In Fig.~\ref{3dpps}, which directly compares the DPPs implicit in the first three figures, the large attractive effect near  3.5 fm due to the excitation of \nuc{16}{O} is very apparent.   From Fig. 2 of Ref.~\cite{OH1}, the effect of the coupling to states of \nuc{16}{O} is mainly to  fill in the deep minimum around $55^\circ$. Fig.~\ref{3dpps} also presents the very small DPP due to coupling to just the $3^-$ state of \nuc{12}{C}. The DPP for the excitation of just the $2^+$ state of \nuc{12}{C} (not shown) is slightly smoother than that when both states of \nuc{12}{C} are  excited, with no direct coupling between them, as is clear from Fig.~\ref{nmc}.

Insight into the contribution of the excitation of states of \nuc{16}{O} may be found by comparing the elastic channel $S$-matrix for the cases with and without these excitations: $S_L$ without the excitation of \nuc{16}{O}  is shown in Fig.~\ref{SL60} and with that excitation in Fig.~\ref{sm0full}.  The different behavior  of $|S_L|$ for the bare potential (dashed lines) directly reflects the lesser depth and radial extent of the bare potential in the latter case. This corresponds to the difference in reaction cross sections when there is no coupling switched on: it is 1633.6 mb for the bare potential used when only the states of \nuc{12}{C} are excited and 1411.7 mb for the bare potential used in the calculations when the states of \nuc{16}{O} are also excited. Also, when comparing these figures, we see a much larger decrease in $|S_L|$ when \nuc{16}{O} coupling is included, and this is reflected in  the increase in reaction cross section due to the coupling. This jumps from 15.6 mb when the states of \nuc{12}{C}
are coupled to 70.4 mb when the states of \nuc{16}{O} are also included (see lines 3 and 6 of Table~\ref{dpps}).  What is somewhat  surprising is the relatively small increase in $\Delta J_{\rm I}$.  Since the bare imaginary potential is somewhat shallower and less diffuse in the full calculation, as is consistent with the behaviour of $|S_L|$ as $L$ approaches 60, it might be expected that the imaginary DPP would compensate for this. Indeed, the reduction in $|S_L|$ directly due to the coupling is greater for the full calculation, but the relatively small increase in $\Delta J_{\rm I}$ 
suggests that this is not entirely due to the imaginary DPP. The extra attraction around 3.5 fm  may contribute  to the absorption by attracting the nuclei into the absorptive potential.  The reduction in $\arg{S_L}$ between $L= 35$ and $L=40$  that can be seen in Fig.~\ref{sm0full} but absent in Fig.~\ref{SL60} suggests  a net repulsive effect in the surface which is not evident in Fig~\ref{full} or Fig.~\ref{3dpps}. 

\begin{figure} [tbh]
\includegraphics[keepaspectratio,width=7.0cm] {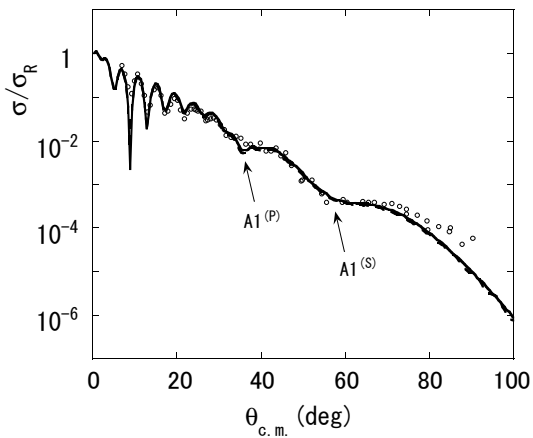}
 \protect\caption{\label{AD} {
Elastic scattering angular distribution for 330 MeV \nuc{16}{O} on \nuc{12}{C}. The circles represent the experimental measurements~\cite{deman},  the solid line is the coupled channel result and the dotted line is for a single channel calculation with the sum of the bare potential and the full DPP. The primary and secondary Airy minima, A1$^{\rm P}$ and A1$^{\rm s}$, are indicated by arrows. } }
\end{figure}

In Fig.~\ref{AD}  we present the elastic scattering angular distributions for the full coupled channel calculation and for single channel scattering due to the sum of the bare potential and the full DPP. They are evidently barely distinguishable. The experimental differential cross sections are included  for comparison, indicating the primary and secondary Airy minima A1$^{\rm P}$ and A1$^{\rm S}$~\cite{OH1}.

\section{DISCUSSION}\label{disco}
It is often very instructive to examine the change in $S_L$, noting that $|S_L|$ is particularly related to the imaginary potential and $\arg{S_L}$ to the real part. Comparing Fig.~\ref{SL60}, which shows the effect of the \nuc{12}{C} states alone, with  Fig.~\ref{sm0full}, which shows the additional effect of the \nuc{16}{O} states, it appears somewhat surprising that it is the  excitation of the \nuc{12}{C} states that has the major effect on correcting the elastic scattering angular distribution.  It should be noted, comparing the dashed lines in these figures, that the significantly greater diffusivity (0.75 fm compared with 0.6 fm) and depth of  the bare imaginary potential for the case without the \nuc{16}{O} excitation has greatly increased $1 - |S_L|$ for $L$ between 40 and 60. The excitation of \nuc{16}{O} has a particularly large effect for this range of partial waves. 

\begin{figure} [tbh]
\includegraphics[keepaspectratio,width=7.0cm] {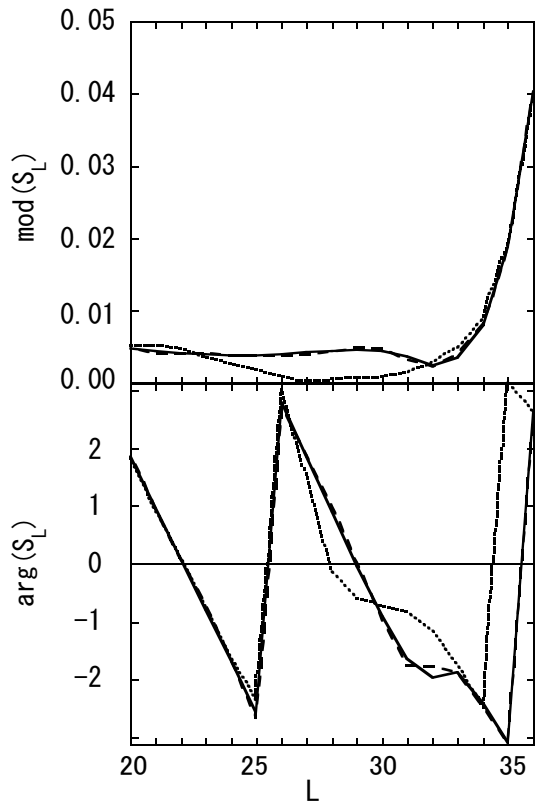}
 \protect\caption{\label{irreg} {
 For $L$ between 20 and 35, $S_L$ for the full calculation with $|S_L|$ in the 
upper panel and $\arg S_L$ in the lower panel. The dotted lines are for an earlier
 stage in the iterative inversion than the dashed lines. The solid lines give the 
CC S-matrix to be inverted the `target' S-matrix.  The final converged potential
 gives $S_L$ that is indistinguishable from the target $S_L$ at this scale.}
} \end{figure}

The DPP generated by the excitation of just the states of \nuc{12}{C} shows that strongly emissive regions can be generated in nuclear interactions by channel coupling. These do not, of course, lead to the breaking of the unitarity limit $|S_L| \le 1$. Such regions  sometimes appear in model independent fits in which the imaginary potential, over particular radial ranges,  becomes small in magnitude or even positive, see for example Ref.~\cite{ermer}. It is likely that such emissivity indicates non-locality and/or $L$-dependence of the underlying DPP. It is interesting that the coupling to \nuc{16}{O} states almost removes the emissive region in this particular case, but nevertheless the potentiality for collective coupling to generate emissivity is clear.

An example that  gives some insight into the very un-smooth shape of the resulting potential, 
and also shows what the inversion process must achieve for the full case, is shown in Fig.~\ref{irreg}. This figure gives a close-up view of $S_L$ for $L$ from 20 to 36. The solid lines show the $S_L$ to be fitted, and the dotted lines show $S_L$ for an early stage of the iterative inversion process. The dashed lines correspond to a substantial change in the potential. For the final potential, the $S_L$ would be indistinguishable from the solid line. The irregular form of $S_L$ in this $L$-region is presumably the result of interference between  amplitudes the origin of which deserves further study.

There is one respect in which the present case conforms to expectations: the last two columns 
of Table~\ref{dpps} indicate that the increase in reaction cross section induced by the collective coupling exceeds the inelastic cross section. This behaviour is not guaranteed and there are cases, e.g.~\cite{pang}, where inelastic processes increase the total cross section by much less than the magnitude of the inelastic cross section.  That is even when the total cross section includes the additional inelastic cross sections. In the present case,  when the coupling to the excited states of \nuc{16}{O} is included, the reaction cross section exceeds the cross section without coupling by substantially more than the total actual inelastic cross section to  all the excited states; compare the difference between 15.6 mb and 14.8 mb on line 3 of Table~\ref{dpps} with that between 70.4 mb and 57.4 mb in line 6. The large value of $\Delta$ CS might be a result of the attractive effect around 3.5 fm drawing the projectile flux into the absorptive region. Recall that the bare potential for the full calculation has a significantly different imaginary term. 

The excitation of just the collective $2^+$ and $3^-$ states of \nuc{12}{C}, both with and without direct coupling between those states, generated rather small overall attraction or repulsion in the surface region, but did generate a deep absorptive feature in the imaginary potential together with a counter-intuitive emissive region at a small separation distance. This, together with the inexact additivity of the local DPPs due to the   $2^+$ and $3^-$ states of \nuc{12}{C} without direct coupling between them, is strongly suggestive of dynamical non-locality of the underlying DPP.   

By contrast with the contribution of states of \nuc{12}{C}, for the case with full coupling the volume integral of the real potential was increased by  about 10\%, the largest increase being around the radius where the bare potential was about half the maximum depth. In the full case, the emissive region at a small separation distance almost completely disappeared.

It is probably true that for many nucleus-nucleus combinations and energies, the radial 
range over which the potential can be determined by elastic scattering data is much less than
for the present case.  For this reason, many discussions of the DPP due to projectile breakup, as calculated  with the continuum discretized coupled-channel method~\cite{rawit74,farell76,yahiro82,austern87,Sakuragi-PTPS-1986}, report the DPP mostly in the surface region.  In fact, even  \nuc{6}{Li} scattering is not that simple~\cite{pang} and in that case, and quite generally for lighter `heavy ions', there is considerable nuclear overlap. As a result, the effect of coupling on the nuclear interaction extends over a radial range where the DPP may have a complicated form, with the effect on the real potential being very different from a uniform renormalization. We conclude from this, and also from the present calculations, that there are situations where the quality and angular range of the experimental data make the uniform renormalization of a folding model or global model potential an inappropriate phenomenological procedure. Correcting a folding model potential with a uniform normalizing factor can make sense in cases where only the surface region is relevant, but otherwise must be considered suspect.   An appropriate phenomenology to exploit precise and wide angular range data would be to add a parameterized model independent correction to a global  optical potential or a folding model potential. This has been done, for example by Khoa \etal~\cite{khoa}. 

The details of the interaction between arbitrary pairs of nuclei are  beyond the reach of global 
models, since they will depend upon specific properties, such as the collectivities, of the particular interacting nuclei.  The procedure employed in the present calculation provides a means of incorporating the particular characteristics of nuclei into a description of their elastic scattering, thus providing a means of going beyond the global models. We comment that Ref.~\cite{pang} reveals the deficiency of the weighted trivially equivalent local potential, TELP, which has sometimes been employed as a means of calculating the DPP.
 
\section{SUMMARY}\label{summa}
The discovery and explanation of a secondary rainbow in \nuc{16}{O}  on \nuc{12}{C} elastic scattering
around 300 MeV provides a conclusive example of the important contribution
made by collective excitations to elastic scattering between nuclei. Specifically, Ref.~\cite{OH1}  explained the occurrence of a secondary rainbow in the elastic scattering differential cross section. The scattering of  \nuc{16}{O}  on \nuc{12}{C} at 330 MeV is sensitive to the internuclear interaction well into the overlap region, affording an opportunity to study the DPP in a situation where little is known about it. Heretofore, DPPs have been evaluated in particular restricted cases:  where one 
projectile is a light nucleus,  where there is no coupling model that is justified by fitting data, 
or where  the more limited TELP inversion procedure is applied. Here, we have applied
IP $S_L \rightarrow V(r)$ inversion in a case where the excitation model is validated 
by its fit to scattering data.  This reveals that the collective excitations generate 
a complicated DPP with both real and imaginary parts having radial forms that depart
very far the bare potentials with uniform multiplicative factors. That is, for the real 
DPP in particular, the effects of the coupling could not be represented by a uniform 
renormalization of the folding model potential. It is reasonable to assume that this 
is a general property of collective contributions in strongly-coupled nucleus-nucleus collisions. 
By studying the DPPs for different combinations of coupled states we have found evidence for the dynamical non-locality of the underlying DPP.

At present, a systematic understanding of the way in which collective excitations, 
and other channel coupling processes, modify the interaction between nuclei, is still lacking. 
This is particularly true regarding the DPP where the interacting nuclei substantially overlap;
here, little is known apart from some cases involving light projectiles, $A \le 6$. 
Such dynamically generated interactions, and the manner in which they depend upon the particular properties of the interacting nuclei,  are accessible by combining coupled channels calculations with $S$-matrix inversion, as we have demonstrated here for strongly overlapping composite nuclei.

\section{ACNOWLEDGEMENTS}
Two of the authors (S.O. and Y.H.) would like to thank the Yukawa Institute for Theoretical Physics for the hospitality extended  during a stay in   2014. 
Part of this work was supported by the Grant-in-Aid for the Global COE Program ``The Next
 Generation of Physics, Spun from Universality and Emergence'' from the Ministry 
of Education, Culture, Sports, Science and Technology (MEXT) of Japan.

\end{document}